\documentstyle[12pt,epsf]{article}
\newcommand{\be}{\begin{equation}}
\newcommand{\ee}{\end{equation}}
\newcommand{\bear}{\begin{eqnarray}}
\newcommand{\ear}{\end{eqnarray}}
\begin{document}
\begin{titlepage}
\begin{flushright}
HD-THEP-97-48\\
hep-ph/yymmnn
\end{flushright}
\vspace{1.5cm}

\begin{center}
{\LARGE Exclusive Hadronic $B$-Decays\footnote{Plenary
talk presented at the b20 Symposium: Twenty Beautiful
Years of Bottom Physics, Chicago, June 29-July 2, 1997}}

\vspace{2cm}
{\large Berthold Stech}\\

\vspace{0.3cm}
Institut f\"ur Theoretische Physik, Universit\"at Heidelberg\\
Philosophenweg 16, D-69120 Heidelberg, Germany\\

\vspace{1.4cm}

{\bf Abstract}\\
\parbox[t]{12cm}{\small  Exclusive non-leptonic two-body decays are
discussed on the basis of a generalized factorization approach
which also includes non-factorizeable contributions. Numerous
decay processes can be described satisfactorily. The success of
the method makes possible the determination of decay constants from
non-leptonic decays. In particular, we obtain $f_{D_s}=(234\pm25)$
MeV and $f_{D^*_s}=(271\pm33)$ MeV. The observed constructive
and destructive
interference pattern in charged $B$- and $D$-decays, respectively,
can be understood in terms of
the different $\alpha_s$-values governing the interaction among
the quarks. The running of $\alpha_s$ is also the cause
of the observed strong increase of the amplitude of lowest
isospin when going to low energy transitions. }

\end{center}

\end{titlepage}

\section{Introduction}
Since we celebrate today 20 years of beauty physics it may be
appropriate to start the discussion of hadronic weak interactions
by briefly recalling what was known about this subject in the seventies.
In spite of many years of intense research on $K$- and hyperon decays,
there was no coherent understanding of non-leptonic decays. For
example, the empirically found dominance of $|\Delta \vec I|=1/2$
transitions over $|\Delta \vec I|=3/2$ transitions by a factor 500 was
a complete mystery. Moreover, the strongest of all
weak decay amplitudes - the $K\to 2\pi$ amplitude -
was found to have to vanish in the $SU3$ symmetry limit (Gell-Mann's
theorem) and no close relation between $K$-decays and hyperon decays
could be seen. In 1974 an important step forward was made:
the construction of an effective Hamiltonian which incorporates
the effects of hard gluon exchange processes\cite{one}. Still, a factor
20 out of the factor 500 could not be explained, nor could
the specific pattern of hyperon
decays. The physics at this time
dealing with $u,d$ and $s$-quarks was not rich enough. In the
corresponding decay processes too few fundamentally different
decay channels are open.

The discovery of open charm in 1976 brought hope for enlightenment.
Many decay channels could now be studied. But also new puzzles
showed up. Unexpectedly, the non-leptonic widths of $D^0$ and $D^+$
turned out to differ by a factor 3 and a strong destructive
amplitude interference in exclusive decays was found. While
$D$-decays occur in a resonance region of the final
particles which complicates the analysis, the discovery of beauty
precisely 20 years ago gave us particles -- the $B$-mesons --
which are ideally suited for the study of non-leptonic decays.
Again, new interesting effects showed up, in particular and
contrary to the case in $D$-decays, a constructive amplitude
interference in charged $B$-decays. Recent results\cite{two} of large
Penguin-type contributions and sizeable transitions to the $\eta'$
particle have still to be understood. Moreover,
$B$-meson decays give
the first realistic possibility to find CP-violating
effects outside the $K$-system.

The dramatic effects observed in hadronic weak decays gave rise
to many speculations. It was a great challenge to find the correct
explanation. Today we know that the strong confining colour forces
among the quarks are the decisive factor. These forces are enormously
effective in low energy processes and still sizeable even in energetic
$B$-decays. Although a strict theoretical treatment of the intricate
interplay of weak and strong forces is not yet possible, a
semi-quantitative understanding of exclusive two-body decays
from $K$-decays to $D$- and $B$-decays has been achieved. The consequences
of the QCD-modified weak Hamiltonian can be explored by
relating the complicated matrix elements of 4-quark operators to
better known objects, to form factors and decay constants.

\newpage
In the present talk I will describe the generalized factorization
method developed recently\cite{three},
which also takes non-factorizeable contributions into
account and has been quite successful so far. It allows the prediction
of many exclusive $B$-decays. I will also show that the
interesting and so far puzzling pattern of amplitude interference in
$B$-, $D$- and $K$-decays is caused by the different values
of $\alpha_s$ acting in these cases.

\section{The effective Hamiltonian}
At the tree level non-leptonic weak decays are mediated by single
$W$-exchange. Hard gluon exchange between the quarks
can be accounted for by using the renormalization group
technique. One obtains an effective Hamiltonian incorporating
gluon exchange processes down to a scale $\mu$ of the order
of the heavy quark mass. For the case of $b\to c\bar ud$ transitions,
e.g., the effective Hamiltonian is
\be\label{1}
H_{eff}=\frac{G_F}{\sqrt 2}V_{cb}V^\star_{ud}\left\{
c_1(\mu)(\bar du)(\bar cb)+c_2(\mu)(\bar cu)(\bar db)
\right\}\ee
where $(\bar du)=(\bar d\gamma^\mu(1-\gamma_5)u)$ etc. are
left-handed, colour singlet quark currents. $c_1(\mu)$
and $c_2(\mu)$ are scale-dependent QCD coefficients known up to
next-to-leading order\cite {four}. Depending on the process considered,
specific forms of the four-quark operators in the effective
Hamiltonian can be adopted. Using Fierz identities one
can put together those quark fields which match the constituents
of one of the hadrons in the final state of the decay process.
Let us consider, as an example, the decays $B\to D\pi$. The
corresponding amplitudes are -- apart from a common factor --
\bear\label{2}
{\cal A}_{\bar B^0\to D^+\pi^-}&=&(c_1+\frac{c_2}{N_c})\langle
D^+\pi^-|(\bar du)(\bar cb)|\bar B^0\rangle,\nonumber\\
&&+c_2\langle D^+\pi^-|\frac{1}{2}(\bar d t^au)(\bar ct^ab)|\bar B^0\rangle
\nonumber\\
{\cal A}_{\bar B^0\to D^0\pi^0}&=&(c_2+\frac{c_1}{N_c})\langle
D^0\pi^0|(\bar cu)(\bar db)|\bar B^0\rangle\nonumber\\
&&+c_1\langle D^0\pi^0|\frac{1}{2}(\bar c t^au)(\bar dt^ab)|\bar B^0\rangle
\nonumber\\
{\cal A}_{B^-\to D^0\pi^-}&=&{\cal A}_{\bar B^0\to D^+\pi^-}
-\sqrt2 {\cal A}_{\bar B^0\to D^0\pi^0}\quad.\ear
$N_c$ denotes the number of quark colours and $t^a$ the Gell-Mann
colour $SU(3)$ matrices. The last relation in (\ref{2})
follows from isospin symmetry of the strong interactions.
The three classes of decays illustrated in eq. (\ref{2}) are referred
to as class I, class II, and class III respectively.

\section{Generalized Factorization}

How shall we deal with the complicated and scale-dependent
four-quark operators?
Because the $(\bar du)$ and the $(\bar cu)$ currents
in (\ref{2}) can generate the $\pi^-$ and $D^0$ mesons,
respectively, the above amplitudes contain the scale-independent
factorizeable parts
\bear\label{3}
{\cal F}_{(\bar BD)\pi}&=&\langle \pi^-|(\bar du)|0\rangle \langle D^+|(\bar cb)|\bar B^0\rangle,
\nonumber\\
{\cal F}_{(\bar B\pi)D}&=&\langle D^0|(\bar cu)|0\rangle \langle \pi^0|(\bar db)|\bar B^0\rangle
\ear
which can be expressed in terms of the decay constants
$f_\pi$ and $f_D$, and the single current transition form
factors $B\to D$ and $B\to\pi$, respectively. For the non-factorizeable
contributions we define hadronic parameters $\epsilon_1(\mu)$ and
$\epsilon_8(\mu)$ such that the amplitudes (\ref{2}) take the
form\cite{five,three}
\bear\label{4}
&&{\cal A}_{\bar B^0\to D^+\pi^-}=a_1{\cal F}_{(BD)\pi}\nonumber\\
&&{\cal A}_{\bar B^0\to D^0\pi^0}=a_2{\cal F}_{(B\pi)D}\nonumber\\
&&a_1=(c_1(\mu)+\frac{c_2(\mu)}{N_c})(1+\epsilon_1^{(BD)\pi}
(\mu))+c_2(\mu)\epsilon_8^{(BD)\pi}\nonumber\\
&&a_2=(c_2(\mu)+\frac{c_1(\mu)}{N_c})(1+\epsilon_1^{(B\pi)D}
(\mu))+c_1(\mu)\epsilon_8^{(B\pi)D}\;.
\ear
The effective coefficients $a_1$ and $a_2$ are scale-independent.
$\epsilon_1$ and $\epsilon_8$ obey renormalization-group
equations and their scale dependence compensates the
scale dependence of the QCD coefficients $c_1$ and $c_2$ \cite{three}.
$a_1$ and $a_2$ are process-dependent quantities because
of the process dependence of the hadronic parameters $\epsilon_1$
and $\epsilon_8$. So far, then, Eq. (\ref{4}) provides
a parametrization of the amplitudes only and allows no predictions
to be made.
To get predictions, non-trivial properties of QCD have to be
taken into account. We employ at this point the $1/N_c$ expansion of QCD.
The large $N_c$ counting rules tell us that
$\epsilon_1=O(1/N_c^2)$ and $\epsilon_8=O(1/N_c)$.
Thus one obtains for $a_1$ and $a_2$ in (\ref{4})
\bear\label{5}
a_1&=&c_1(\mu)+c_2(\mu)(\frac{1}{N_c}+\epsilon_8^{(BD)\pi}
(\mu))+O(1/N_c^2)\nonumber\\
a_2&=&c_2(\mu)+c_1(\mu)(\frac{1}{N_c}+\epsilon_8^{(B\pi)D}
(\mu))+O(1/N_c^2)\quad.\ear
For $B$-decays using $c_1(m_b)=1+O(1/N^2_c)$ and $c_2(m_b)=O(1/N_c)$
one finally gets\cite{three}
\newpage
\bear\label{6}
&&a_1=c_1(m_b)+O(1/N_c^2)\nonumber\\
&&a_2=c_2(m_b)+\zeta^Bc_1(m_b)+O(1/N_c^3)\ear
with
\[c_1(m_b)\approx1\quad {\rm and}\quad \zeta^B=\frac{1}{N_c}
+\epsilon_8^{(B\pi)D}(m_b)\quad.\]
Now, neglecting $O(1/N^2_c)$ terms, we are left with a single
parameter $(\zeta^B)$ only. It should be emphasized that putting
this parameter equal to $1/N_c$ does not correspond to any consistent
limit of QCD. For $a_2$ the more general expression (\ref{6}) 
must be used\cite{7,six}.

$\zeta^B$ is a dynamical parameter: In general, it will take different
values for different decay channels. To deal with this, let us
introduce a process-dependent factorization scale $\mu_f$
defined by $\epsilon_8(\mu_f)=0$. The renormalization-group
equation then gives\cite{three}
\be\label{7}
\epsilon_8(\mu)=-\frac{4\alpha_s}{3\pi}\ln \frac{\mu}{\mu_f}
+O(\alpha^2_s)\quad.\ee
For different processes the variation of the factorization
scale $\mu_f$ is expected to scale with the energy release
to the outgoing hadrons in the decay process. With
$\mu_f\approx O(m_b)$ one gets from (\ref{6}), \ (\ref{7})
\be\label{8}
\Delta\zeta^B\approx\frac{4\alpha_s}{3\pi}\frac{\Delta\mu_f}{m_b}
\approx \ {\rm few}\ \%\quad.\ee
Thus, the process dependence of $\zeta^B$ is expected to be
very mild. To a good approximation a single value
appears sufficient for the description of two-body $B$-decays.
One finds (see section 4) $\zeta^B=0.45\pm0.05$.

A similar discussion also holds for $D$-decays. There
one  is led to\cite{three}
\bear\label{9}
    a_1&\approx& c_1(m_c)+\zeta^{'D}c_2(m_c)\nonumber\\
    a_2&\approx& c_2(m_c)+\zeta^Dc_1(m_c)\nonumber\\
\zeta^{'D}&\approx& \zeta^D \ear
and again expects only a mild process dependence of $\zeta^D$.
Indeed, the corresponding description of exclusive $D$-decays
brought reasonable success. $\zeta^D$ turned out to be
very small or zero. There is also theoretical support
(using QCD sum rule methods) for a 
partial or full cancellation of the $1/N_c$ term by non-factorizeable
contributions\cite {seven}.
On the other hand, the
corresponding calculation of $\zeta^B$ is more involved\cite{eight}
and was so far not successful.

\section{Determination of \boldmath $\lowercase {a_1}$ and
\boldmath $\lowercase{a_2}$}
The most direct way to determine the effective constant $a_1$
consists in comparing non-leptonic decay rates with
the corresponding differential semi-leptonic rates at
momentum transfers equal to the masses of the current
generated particles\cite{nine}. One gets, for example,
\be\label{10}
\frac{\Gamma(\bar B^0\to D^{(*)+}\rho^-)}{d\Gamma
(\bar B^0\to D^{(*)+}\ell^-\bar\nu)/dq^2
|_{q^2=m^2_\rho}}=6\pi^2|V_{ud}|^2f^2_\rho|a_1|^2\quad.\ee
Because the generated particle is a vector particle
like the lepton pair, the form factor combinations occurring
in the nominator and denominator cancel precisely. Thus,
the ratio (\ref{10}) is solely determined by $|a_1|$
and the $\rho$-meson decay constant $f_\rho$.
Taking by convention $a_1$ real and positive, the measured
rates\cite{ten} give\cite{three}
$a_1=1.09\pm0.13$
in agreement with the expectation (\ref{6}). $a_1$ values
obtained from several other processes are in full agreement
with the above number. In transition to pseudoscalar particles
the form factor combinations in equations replacing (\ref{10})
do not cancel. But for $B\to D,D^*$ matrix elements all
form factors are well determined using experimental data and
the heavy quark effective theory\cite{eleven}. The latter relates
in particular longitudinal form factors to the
transversal ones.

Values for $|a_2|$ can be obtained from the analysis of class II
transitions. The decays $\bar B^0\to D^{0(*)}h^0$
$(h^0:\pi^0, \rho^0, a^0_1)$ have not yet been observed, but the
branching ratios for $\bar B\to K^{(*)} J/\psi$ and
$\bar B\to K^{(*)}\psi(2S)$ are available\cite{ten}. The analysis
requires model estimates for the heavy-to-light form factors,
which enter here. We use the NRSX model\cite{twelve} which is
based on the extrapolation of the BSW form factors\cite{six}
at $q^2=0$ by appropriate pole and dipole formulae.
Where available, more sophisticated calculations agree with
these results. (See. e.g. Ref. 14). We find\cite{three}
$|a_2|=0.21\pm0.01\pm0.04$, where the second error accounts
for the model dependence.

The relative phase between $a_2$ and $a_1$ together with
the magnitude of $a_2$ can be obtained from the decays
$B^-\to D^{(*)0}h^-$ where, as seen from (\ref{2}) and
(\ref{4}), the two amplitudes interfere.
The data for the ratios $\Gamma(B^-\to D^{(*)0}h^-)
/\Gamma(\bar B^0\to D^{(*)+}h^-)$ give conclusive
evidence for constructive interference\cite{ten}. Taking
$a_2$ to be a real number (vanishing final state interaction),
we find\cite{three} $a_2/a_1=+0.21\pm0.05\pm0.04$.
Combined with the value for $a_1$ this gives $a_2=+0.23\pm0.05\pm0.04$.
The nice agreement between the two determinations of $|a_2|$ shows
that the process dependence of this quantity cannot be
large. There is no evidence for it. An analysis with an alternative
and very simple form factor model gives slightly larger
values for $a_2$ but the results from different processes are
again consistent with each other\cite{three}.

The positive value for $a_2/a_1$ in exclusive $B$-decays
is remarkable. It is different from the value
of the same
ratio in exclusive $D$-decays. There $a_2/a_1$ is negative
causing a sizeable destructive amplitude interference. The change
of $a_2/a_1$ by
going from $B$- to $D$-and $K$-decays will be discussed
in section 6.

\section{Tests and Results}

The $B$-meson, because of its large mass, has many decay
channels. We learned from important examples the values
of $a_1$ and $a_2$ and their near process-independence in
energetic two-body decays. Thus numerous tests and
predictions for branching ratios and for the polarizations
of the outgoing particles can be made. I will be very brief
here and simply refer to Ref. 3 for the compilation of
branching ratios in tables, for a detailed discussion and
for comparison with the data. Also discussed there is the
possible influence of final state interactions. Limits 
on the relative phases of isospin amplitudes are given. 
In contrast to $D$-decays final state interactions do not 
seem to play an important role for the dominant 
exclusive $B$-decay modes.
For the much weaker Penguin-induced transitions, 
$\bar B\to K^{(*)}\pi$ for example, this statement 
does not hold. Small amplitudes can get an additional 
contribution from stronger decay 
channels\cite{six,fourteen}. In the $\bar B\to
K^{(*)}\pi$ case the decay can proceed via virtual 
intermediate $D^{(*)}\bar D^{(*)}_s$ like channels 
generated by the $b\to c\bar cs$
interaction. The colour octet $c\bar c$ pair, if at low
invariant mass, may then turn into a pair of light 
quarks by gluon exchange. This gives rise to a 
"long range Penguin" contribution\cite{fourteen} in 
addition to the short distance Penguin amplitude.
In future application of our generalized factorization 
method
to rare decays this should be kept in mind. Here, 
however, I will not discuss this subject further.

Non-leptonic decays to two spin-1 particles also need a
separate discussion. Here one has 3 invariant amplitudes
corresponding to
outgoing $S$, $P$, and $D$-waves. Non-factorizeable
contributions to these amplitudes may, in general,
have an amplitude composition different from the factorizeable
one which cannot be dealt by introducing effective $a_1$ parameters.
Whether or not and to what
extent factorization also holds in these more complicated
circumstances can be learned from the polarization of the final
particles. In class I decays the factorization approximation
predicts a polarization identical to
the one occurring in the corresponding
semi-leptonic decays at the appropriate $q^2$ value.
For $B\to D^*V$ decays the theoretical predictions
have very small errors only\cite{three}. Another case of particular
interest is the
polarization of the $J/\psi$ particle in the decay
$B\to K^*\ J/\psi$. Form factor models predict
a longitudinal polarization of around 40\% .
A recent CLEO measurement\cite{fifteen} gives $(52\pm7\pm4)\%$ .
It can be shown\cite{sixteen} that small changes of the ratios
of form factors obtained in the NRSX model at $q^2=m^2_{J/\psi}$
are sufficient to get full agreement with the measurements
of the longitudinal as well as both
transverse polarizations. At present,
even with respect to polarization measurements, the generalized
factorization approximation is in agreement with
the data.

Because of its success, the generalized factorization method,
besides allowing many predictions for yet unmeasured decays,
can also be used to determine unknown decay constants.
A case in point is the determination of the decay constant
of the $D_s$ and $D^*_s$ particles.
Comparing non-leptonic decays to $D_s, D^*_s$ with those to
light mesons, we find\cite{three}
\be\label{11}
f_{D_s}=(234\pm25)\ {\rm MeV},\quad f_{D^*_s}=
(271\pm33)\ {\rm MeV}\;.\ee
In this determination $a_1$ cancels and, presumably, also
some of the experimental systematic errors. The value for
$f_{D_s}$ is in excellent agreement with the value $f_{D_s}
=(241\pm37)$ MeV obtained from the leptonic decay of the
$D_s$ meson\cite{seventeen}. There are several other decay constants
which can be measured this way. Of particular interest are the
decay constants of $P$-wave mesons like the
$a_0,a_1,K^*_0,K_1$ particles.

\section{From \boldmath $B$- to \boldmath $D$- to \boldmath $K$-Decays}

The process dependence of the coefficients $a_1$ and $a_2$
governing exclusive $B$-decays turned out to be very mild.
In fact, it is not seen within the errors of the present data.
But $a_1$ and $a_2$ change strongly by going from $B$-decays
to $D$-decays or even down to $K$-decays. In the generalized
factorization scheme this is expected because of the different
factorization scales and the corresponding $\alpha_s(\mu_f)$
values controlling the strength of the colour forces between
the quarks. In Fig. 1
the ratio $a_2/a_1$ is plotted as a function of
$\alpha_s(\mu_f)$ . We used for the Wilson coefficients
the renormalization group invariant definitions of Ref. 19.
It appears appropriate for describing the changes of
the scale-independent coefficients $a_1$ and $a_2$ with changing
the particle energy. As seen from the figure the positive
value of $a_2/a_1$ found for exclusive
$B$-decays indicates that here small values of $\alpha_s$ govern
the colour forces in the first instant of the decay process. This
is an impressive manifestation of the colour
transparency argument put forward by Bjorken\cite{nine}.
In $D$-decays the stronger gluon interactions redistribute
the quarks: the induced neutral current interaction is already
sizeable. We took the corresponding values of $a_1$ and $a_2$
from the measured isospin amplitudes. They are less affected
by final state interactions than the individual amplitudes.
The ratio $|A_{1/2}|/A_{3/2}|$ is already rather large
$(\approx 4)$ leading to $a_2/a_1 \approx -0.45$ . According
to the figure this corresponds to an effective value
$\alpha_s \approx 0.7$. The negative value of $a_2$,
and the corresponding destructive amplitude
interference in charged $D$-decays, has been known for many
years\cite{six,nineteen}. Since
the bulk of $D$-decays are two-body or quasi two-body decays,
it is the main cause for the lifetime difference of $D^+$
and $D^0$ in full accord with estimates of the relevant partial
inclusive decay rates\cite{twenty}.

\begin{figure}
\epsfxsize=7cm
\centerline{\epsffile{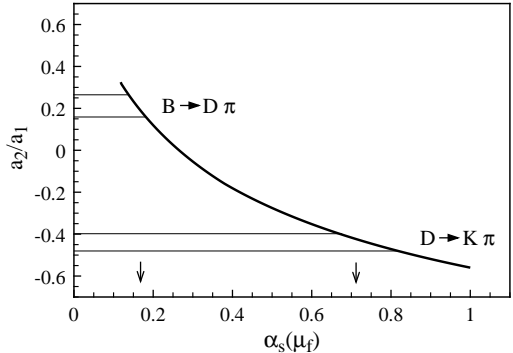}}
\caption{\label{Fig:a2a1}
The ratio $a_2/a_1$ as a function of the running coupling constant
evaluated at the factorization scale. The bands indicate the
phenomenological values of $a_2/a_1$ extracted from $\bar B\to D\pi$
and $D\to K\pi$ decays.}
\end{figure}

Because of the onset of non-perturbative effects
one cannot extent Fig. 1 down to
larger $\alpha_s$ values. However, the trend to smaller
and smaller values of
the ratio of the Wilson coefficients $c_+(\mu_f)/c_-(\mu_f)$,
which is already down to
$\approx 0.17$ for $D$-decays, is visible. It indicates
a strong and, presumably, non-perturbative force in the
colour $3^*$ channel of two quarks, i.e. in the scalar diquark
channel\cite{twentyone}. In $K$-decays one is very close to the limiting
case $a_2/a_1=-1$ for which the $|\Delta \vec I|=1/2$ rule
would hold strictly.

\section{Conclusions}

The matrix elements of non-leptonic exclusive decays are
notoriously difficult to calculate. Factorization provides
for a connection with better known objects. If combined
with the $1/N_c$ expansion method and properly
applied and interpreted, it turns out to be very useful,
at least for energetic $B$-decays, and has passed
many tests. Thus it enables reliable predictions for many
decay channels to be made and also permits the determination of decay
constants which are difficult to measure otherwise. Factorization
does not necessarily hold to the same degree for
transitions to two vector particles. These are more sensitive
to non-factorizeable contributions and final state interactions.

The constant $a_1$ is predicted to be one
apart from $1/N^2_c$ corrections in exclusive $B$-decays and to
be practically
process-independent. The analysis confirmed these expectations.
The particularly interesting parameter $a_2$, within
errors, also does not show a process dependence.
The positive value of $a_2/a_1$ extracted from
exclusive $B$-decays is remarkable. The obvious interpretation
is that a fast-moving colour singlet quark pair interacts
little with soft gluons.
The constructive interference in energetic two-body $B^-$-decays
does not imply that the lifetime of the $B^-$-meson should be
shorter than the lifetime of the $\bar B^0$ meson: The majority
of transitions proceed into multi-body final states. For these
the relevant scale may be lower than $m_b$ leading to destructive
interference. Also, there are many decay channels for which
interference cannot occur.
The running of $a_1$ and $a_2$ with $\alpha_s(\mu_f)$, which in
turn depends on the energy release to the final particles,
is very interesting. It causes the change from constructive
amplitude interference in $B^-$-decays to strong destructive
interferences in $D$- and $K$-decays. Since exclusive two-body
and quasi two-body decays are dominant in $D$-decays this
destructive interference is the main cause of the lifetime
difference between $D^0$ and $D^+$. By going to low energies
the lowest isospin amplitude is seen to become more and more dominant.
Strange particle decays are the most spectacular manifestation
of the dramatic changes occuring when the effective $\alpha_s$
gets large. A unified picture of exclusive non-leptonic decays
emerges which ranges from very low scales to the large energy
scales relevant for $B$-decays.

\section{Acknowledgement}
The work reported here was performed in a fruitful and most
enjoyable collaboration with Matthias Neubert which is gratefully
acknowledged. The author also likes to thank Dan Kaplan and the
other organizers of the b20 symposium for the very pleasent meeting
and Matthias Jamin for a useful discussion.


\begin{thebibliography}{99}

\bibitem{one}
G. Altarelli and L. Maiani, , {\it Phys. Lett. B} {\bf 52}
(1974) 351;\\
M. K. Gaillard and B. W. Lee, {\it Phys. Rev. Lett.}
{\bf 33} (1974) 108.
\bibitem{two}
F. W\"urthwein CLEO Collaboration, hep-ex/9706010;\\
R. Poling CLEO Collaboration, these Proceedings.
\bibitem{three}
M. Neubert and B. Stech, hep-ph 9705292, to appear in
{\it Heavy Flavours}, Second
Edition, ed. A. J. Buras and M. Lindner
(World Scientific, Singapore).
\bibitem{four}
G. Altarelli, G. Curci, G. Martinelli, and S. Petrarca,
{\it Phys. Lett. B} {\bf 99} (1981) 141; {\it Nucl. Phys. B}
{\bf 187} (1981) 461;\\
 A. J. Buras and P. H. Weisz, {\it Nucl. Phys. B}
{\bf 333} (1990) 66.
\bibitem{five}
 H. Y. Cheng, {\it Phys. Lett. B} {\bf 335} (1994) 428;\\
 J. M. Soares, {\it Phys. Rev. D} {\bf 51} (1995) 3518.
\bibitem{six}
 M. Bauer and B. Stech, {\it Phys. Lett. B} {\bf 152} (1985) 380;\\
 M. Bauer, B. Stech, and M. Wirbel, {\it Z. Phys. C}
{\bf 34} (1987) 103.
\bibitem{7}
N. Deshpande, M. Gronauand and D. Sutherland, {\it Phys. Lett. B}
{\bf 90} (1980) 431, {\it Nucl. Phys. B} {\bf 183} (1981) 367.
\bibitem{seven}
 B. Blok and M. Shifman, {\it Yad. Fiz.} {\bf 45} (1987)
221, 478, and 841 [{\it Sov. J. Nucl. Phys.}
{\bf 45} (1987) 135, 301, and 522]; {\bf 46} (1987) 1310
[{\bf 46} (1987) 767].
\bibitem{eight}
 B. Blok and M. Shifman, {\it Nucl. Phys. B} {\bf 389}
(1993) 534; {\bf 399} (1993) 441 and 459.
\bibitem{nine}
J. D. Bjorken, {\it Nucl. Physcis B} (Proc. Suppl.) {\bf 11}
(1998) 325.
\bibitem{ten}
T. E. Browder, K. Honscheid, and D. Pedrini, {\it Ann. Rev.
Nucl. Part. Sci.} {\bf 46} (1996) 395.
\bibitem{eleven}
For a review see M. Neubert, {\it Phys. Rep.} {\bf 245} (1994)
259; {\it Int. J. Mod. Phys. A} {\bf 11}
(1996) 4173.
\bibitem{twelve}
M. Neubert, V. Rieckert, B. Stech, and Q. P. Xu, in
{\it Heavy Flavours}, First Edition, ed. A. J. Buras and M. Lindner
(World Scientific, Singapore 1992), p. 286.
\bibitem{thirteen}
P. Ball and V. M. Braun, {\it Phys. Rev. D} {\bf 55} (1997) 5561.
\bibitem{fourteen}
W. F. Palmer and B. Stech, {\it Phys. Rev. D}
{\bf 48} (1993) 4147.
\bibitem{fifteen}
J.D. Lewis, Proceedings, B-Physics and CP Violation Conf.
Honolulu, HI (1997) to be published.
\bibitem{sixteen}
Y.Y. Keum, privat communication.
\bibitem{seventeen}
J. D. Richman, {\it Proceedings of the 28th International
Conference on High-Energy
Physics}, Warsaw, Poland, July 1996 ICHEP 96:143 [hep-ex/9701014].
\bibitem{eighteen}
A.J. Buras, M. Jamin, E. Lauterbacher, and P. Weisz,
{\it Nucl. Phys. B} {\bf 370} (1992) 104, 501.
\bibitem{nineteen}
A. J. Buras, J. M. G\'erard, and R. R\"uckl,
{\it Nucl. Phys. B} {\bf 268} (1986) 16.
\bibitem{twenty}
B. Stech, {\it Nucl. Phys.} (Proc. Suppl.) {\bf B1} (1988) 17;\\
B. Stech, in {\it CP-violation}, ed. C. Jarlskog
(World Scientific, Singapore 1989) p. 680.
\bibitem{twentyone}
M. Neubert and B. Stech {\it Phys. Rev. D} {\bf 44} (1991) 775;\\
B. Stech, {\it Mod. Phys. Lett. A} {\bf 6} (1991) 3113;\\
M. Jamin and A. Pich, {\it Nucl. Phys. B} {\bf 425} (1994) 15.

\end{thebibliography}
\end{document}